\documentclass[twocolumn,showpacs,prc,aps,floats]{revtex4}
\usepackage{graphicx}
\newcommand{\ds }{\displaystyle}
\newcommand{\ra}{\rightarrow}
\newcommand{\be}{\begin{equation}}
\newcommand{\ee}{\end{equation}}
\newcommand{\bea}{\begin{eqnarray}}
\newcommand{\eea}{\end{eqnarray}}
\newcommand{\ci}{\cite}
\newcommand{\bi}{\bibitem}
\newcommand{\nono}{\nonumber \\}

\newcommand{\dd}{\partial}

\newcommand{{\bfna}}{\mbox{\boldmath$\vec{\nabla}$}}

\def\dal{\,\lower0.3ex\vbox{\hrule\hbox{\vrule\kern2pt\vbox{\kern4pt\kern4pt}
\kern2pt\vrule}\hrule}\,}

\begin{document}

\title{ Assisted tunneling of a wave packet between square barriers}
\author{G. K\"albermann}
\email{hope@vms.huji.ac.il}
\affiliation{Soil and Water department, 
The Robert H. Smith Faculty of Agriculture, Food and Environment 
Hebrew University, Rehovot 76100, Israel}

\begin{abstract}

The assisted tunneling of a wave packet between square one dimensional barriers
is treated analytically. The survival probability is calculated exactly for a 
potential mimicking a constant electric field with arbitrary time dependence.
The pole spectrum of the properly normalized unperturbed wavefunctions 
determines the decay time.
The tunneling probability is enhanced by the perturbation.
The results are exemplified for a simplified model of nuclear $\ds \alpha$ 
decay. 
\end{abstract}
\pacs{ 03.65.Xp, 73.43.Jn 23.60.+e}
\maketitle

\section{\sl Introduction}

The quantum tunneling paradigm explains diverse processes,
as currents in Josephson junctions and nuclear alpha decay. 
Energetically forbidden regions for a classical object, become
accessible for a quantum system.

Gurney and Condon and Gamow\ci{condon, gamow}
, used the tunneling model to reproduce 
the experimentally measured huge range of
$\alpha$ decay lifetimes of similar unstable nuclei.
The scheme has withstood continuous scrutiny in the intervening years
\ci{medeiros, holstein}.

In a previous work we investigated the tunneling of a one dimensional 
metastable state between delta function barriers, excited by a time dependent
potential.\ci{g1} By inspecting the modification of the spectrum of the
poles ruling the decay process, we estimated a non-negligible enhancement of
the tunneling probability when
an external time dependent perturbation acts on the system.

In the present work we find exact solutions to the assisted tunneling process,
with square barrier potentials instead of delta functions. 
Square potentials provide a more realistic description of the barriers
involved in decay processes.

Decay times are enormously 
different form the natural time scales of nuclear and even atomic phenomena.
It is not possible to follow the decay process numerically.
Analytical expressions are then of the utmost importance\ci{g1,van,garcia}. 
The method and formulas presented here aim at contributing in this direction.
In the next section we present the model and the analytical solution method.
In section 3 we solve the equations for a time harmonic, linear potential and
evaluate the effect of the perturbation on the tunneling process.
Some mathematical details are presented in appendix A.

\section{\sl Analytical method for assisted tunneling}

The Schr\"odinger equation for a one dimensional system
between square barriers is 

\be\label{h1}
i~\frac{\dd\Psi}{\dd t}=\frac{-1}{2~m}\frac{\dd^2 \Psi}{\dd x^2}+
\gamma~(\Theta(d-|x|)\Theta(|x|-x_0))\Psi
\ee

\noindent In the following all our units are either
{\sl fm} for length and time, and {\sl$\ds fm^{-1}$} for energies, momenta and
potential strength $\ds \gamma$.

We consider an initial Gaussian wavepacket
located $\ds t=0$ in the region between the barriers.
\bea\label{packet}
\Psi(x,t=0)=e^{-\frac{x^2}{\Delta^2}}
\eea
\noindent where $\ds \Delta$ is the width parameter of the packet.
This packet is a reasonable model for the metastable state and 
facilitates the analytical evaluation of the decay amplitudes. However, 
the method presented here is not limited to gaussian packets.

Inside the barrier region the packet disperses if $\ds \Delta~<~x_0$.
, but, soon enough swelling comes to a halt due to the presence 
of the barriers. It can then
spread only through the tunneling process governed by the barriers.

The even $\ds \chi_e(x)=\frac{\varphi_e(k)}{\sqrt{\pi} n_e(k)}$ and odd $\ds \chi_o(x)=\frac{\varphi_o(k)}{\sqrt{\pi} n_o(k)}$ stationary states of eq.(\ref{h1}) for 
energies below the barrier height $\ds \gamma$ are 

\bea\label{even}
\varphi_e(k)=\left \{
\begin {array}{l}
cos(k x);|x|<x_0\\
A_1 e^{\kappa |x|}+  B_1 e^{-\kappa |x|}~;d>|x|>x_0\\
C_1 cos(k x)+sign(x) D_1 sin(k x);~|x|>d
\end{array}\right.
\eea

\noindent where $\ds \kappa$ is defined in eq.(\ref{ne_1}), 
and $\ds sign(x)=1,-1$ for $\ds x>0, x<0$.

\bea\label{odd}
\varphi_o(k)=\left \{
\begin {array}{l}
sin(k x);~|x|<x_0\\
sign(x)( A_2 e^{\kappa |x|}+ B_2 e^{-\kappa |x|});~d>|x|>x_0\\
sign(x) C_2 cos(k x)+ D_2 sin(k x);~|x|>d
\end{array} \right.
\eea
\noindent where we have extracted a factor of $\ds \sqrt{\pi}$ from the 
normalizations for convenience.

The set of even-odd functions is orthonormal and complete.\ci{bron,patil} 
The zeros of the normalization factors 
govern the exponential decay of 
the metastable wave functions in the inner region\ci{g1}. 

The normalization factors are given by

\bea\label{neven}
((n_e(k))^2&=&(C_1(k)^2+D_1(k)^2)\nono
D_1&=&-\frac{1}{2}~\bigg(c_2~\kappa~e_2~k~s_1\nono
&-&c_2~\kappa^2~e_2~c_1+c_2~\kappa^2~e_1~c_1\nono
&+&c_2~\kappa~e_1~k~s_1+e_2~k^2~s_1~s_2-e_2~\kappa~c_1~k~s_2\nono
&-&e_1~\kappa~c_1~k~s_2-e_1 k^2~s_1~s_2\bigg)/(\kappa~k~e_3)\nono
C_1&=&\frac{1}{2}\bigg(e_2~k~s_1~\kappa~s_2-e_2~\nono
&\kappa&^2~c_1~s_2+e_1~\kappa^2~c_1~s_2+e_1~k~s_1~\kappa~s_2\nono
&-&c_2~e_2~k^2~s_1+c_2~e_2~\kappa~c_1~k+c_2~e_1~\kappa~c_1~k\nono
&+&c_2~e_1~k^2~s_1\bigg)/(\kappa~k~e_3)
\eea

\bea\label{nodd}
(n_o(k))^2&=&(C_2(k)^2+D_2(k)^2)\nono
D_2&=&-\frac{1}{2}~\bigg(-c_2~e_2~\kappa~c_1~k\nono
&-&c_2~\kappa^2~e_2~s_1+c_2~\kappa^2~e_1~s_1\nono
&-&c_2~e_1~\kappa~c_1~k-e_2~k^2~c_1~s_2-e_2~k~s_1~\kappa~s_2\nono
&-&e_1~k~s_1~\kappa~s_2+e_1~k^2~c_1~s_2\bigg)/(\kappa~k~e_3)\nono
C_2&=&-\frac{1}{2}\bigg(e_2~\kappa~c_1~k~s_2\nono
&+&e_2~\kappa^2~s_1~s_2-e_1~\kappa^2~s_1~s_2\nono
&+&e_1~\kappa~c_1~k~s_2-c_2~e_2~k^2~c_1-c_2~\kappa~e_2~k~s_1\nono
&-&c_2~\kappa~e_1~k~s_1+c_2~e_1~k^2~c_1\bigg)/\nono
&(&\kappa~k~e_3)\eea

\noindent where
\bea\label{ne_1}
\kappa&=&\sqrt{2~m~\gamma-k^2}\hspace{2.2 true cm}e_3=e^{\kappa~(x_0+d)}\nono
e_2&=&e^{2~\kappa~d}\hspace{3.6 true cm}e_1=e^{2~\kappa~x_0}\nono
c_2&=&cos(k~d)\hspace{3.1 true cm}s_2=sin(k~d)\nono
c_1&=&cos(k~x_0)\hspace{2.9 true cm}s_1=sin(k~x_0)\nono
\eea

We assume a time dependent spatially linear perturbation potential

\bea\label{potential}
V(x,t)=\mu~x~G(t)
\eea

\noindent with $\ds \mu$ a coupling constant. For a spatially constant 
time-harmonic electric field of intensity $\ds E_0$ interacting with 
an $\ds \alpha$ particle of charge $\ds 2~|e|$, we have

\bea\label{th}
~\mu~G(t)=2~|e|~E_0~sin(\omega t)
\eea

In order to eliminate the explicit space
dependence of the perturbation term in the Schr\"odinger equation, we apply the unitary transformation

\bea\label{unitary}
\Psi(x,t)&=&e^{-i\sigma}\Phi\nono
\sigma&=&\mu~x~\zeta+\int{\frac{\zeta^2~\mu^2}{2~m}~dt}\nono
\zeta&=&\int{G(t)~dt}
\eea
\noindent The Schr\"odinger equation(\ref{h1}) including the
perturbation of eq.(\ref{potential}) becomes

\bea\label{h2}
i~\frac{\dd\Phi}{\dd t}&=&\frac{-1}{2~m}\frac{\dd^2 \Phi}{\dd x^2}+
\gamma~(\Theta(d-|x|)\Theta(|x|-x_0))\Phi\nono
&+&i\zeta(t)~\frac{\mu}{m}\frac{\dd \Phi}{\dd x}
\eea

The wave function $\ds \Phi$ 
is expanded in the complete set of even and odd states of the unperturbed 
Schr\"odinger equation (\ref{h1}) including states with energies above the barrier

\bea\label{expansion}
\Phi(x,t)=\sum_{i=e,o}\int_0^{\infty}\chi_i(k,x)~c_i(k,t)~e^{
\frac{-i~k^2~t}{2~m}}~dk
\eea

The amplitudes $\ds c_{e,o}$ consequently obey the time evolution

\bea\label{sch1}
\dot{c}_o(k)&=&-\frac{\mu~k~\zeta}{m~n_e(k)~n_o(k)}~c_e(k)\nono
\dot{c}_e(k)&=&\frac{\mu~k~\zeta}{m~n_e(k)~n_o(k)}~c_o(k)
\eea

\noindent where we have used the superposition 
integral calculated in appendix A
\footnote{In ref.\ci{g1} there was a sign misprint 
in the second of eqs.(11)}

\bea\label{superp}
I&=&\int{e^{-\frac{i(k'^2-k^2) t}{2m}}
\chi_e(k,x)\frac{\dd \chi_o(k',x)}{\dd x} dx}\nono
&=&\frac{k}{n_e(k) n_o(k)}\delta(k-k')
\eea

\noindent and a dot denotes a time derivative.

The exact solution to eqs.(\ref{sch1}) for an initially symmetric wave packet
such as the one of eq.(\ref{packet}) is

\bea\label{ampl}
c_e(k,t)&=&c_e(k,0)~cos(\frac{\mu~k~Y}{m~n_e~n_o})\nono
c_o(k,t)&=&-c_e(k,0)~sin(\frac{\mu~k~Y}{m~n_e~n_o})\nono
Y(t)&=&\int_0^t{\zeta(t')~dt'}\nono
c_e(k,0)&=&\frac{\Delta}{\pi^{\frac{1}{4}}n_e(k)}~e^{-\frac{k^2}{4~\Delta^2}}
\eea

\noindent where $\ds c_e(k,0)$ corresponds a well localized $\ds 
\Delta<<x_0$
 wave packet of the form of eq.(\ref{packet}). Other initial wave
functions merely change the functional dependence of $\ds c_e(k,0)$.

\section{\sl Assisted tunneling solution}

We are interested in the decay of the wave packet from the internal zone 
leaking out to infinity. As discussed in \ci{g1}, the most 
important contribution to the integral in eq.(\ref{expansion}) originates 
from the poles in the amplitudes of eq.(\ref{ampl}).
This is indeed correct for the unperturbed decay process. However, for
the perturbed process, the amplitudes are oscillatory. The poles
lie in extremely close proximity to the real momentum axis. When
the integration over this variable is effected, 
the argument of the harmonic functions in eq.(\ref{ampl}) can oscillate 
wildly, suppressing the contribution of the poles.(see figures 2 and 3 below)
However, if the argument of
the harmonic functions is of order one, the poles will still dominate.
This can be achieved by choosing a sufficiently high frequency for 
the external perturbation of eq.(\ref{th}).
 
The normalization factors around their minima can be cast in the form

\bea\label{pole_1}
k^2~n_{e,o}^2&\approx&~\lambda_{j,(e,o)}~(k^2-{k_{j,(e,o)}}^2)^2
+\beta_{j,(e,o)}
\eea
\noindent where {\sl j} enumerates the zeros of $\ds n_e, n_o$.

The minima are located close to

\bea\label{beta}
k_e&\approx&\frac{(2n+1)~\pi~p}{2~(1+p~x_0)}~\nono
k_o&\approx&\frac{n~\pi~p}{(1+p~~x_0)}\nono
p&=&\sqrt{2~m~\gamma}
\eea
Eqs.(\ref{beta}) provide starting 
values for the numerical search of the location of the poles of the wave 
function.
\begin{figure}[htb]
\includegraphics[width=9cm,height=10cm]{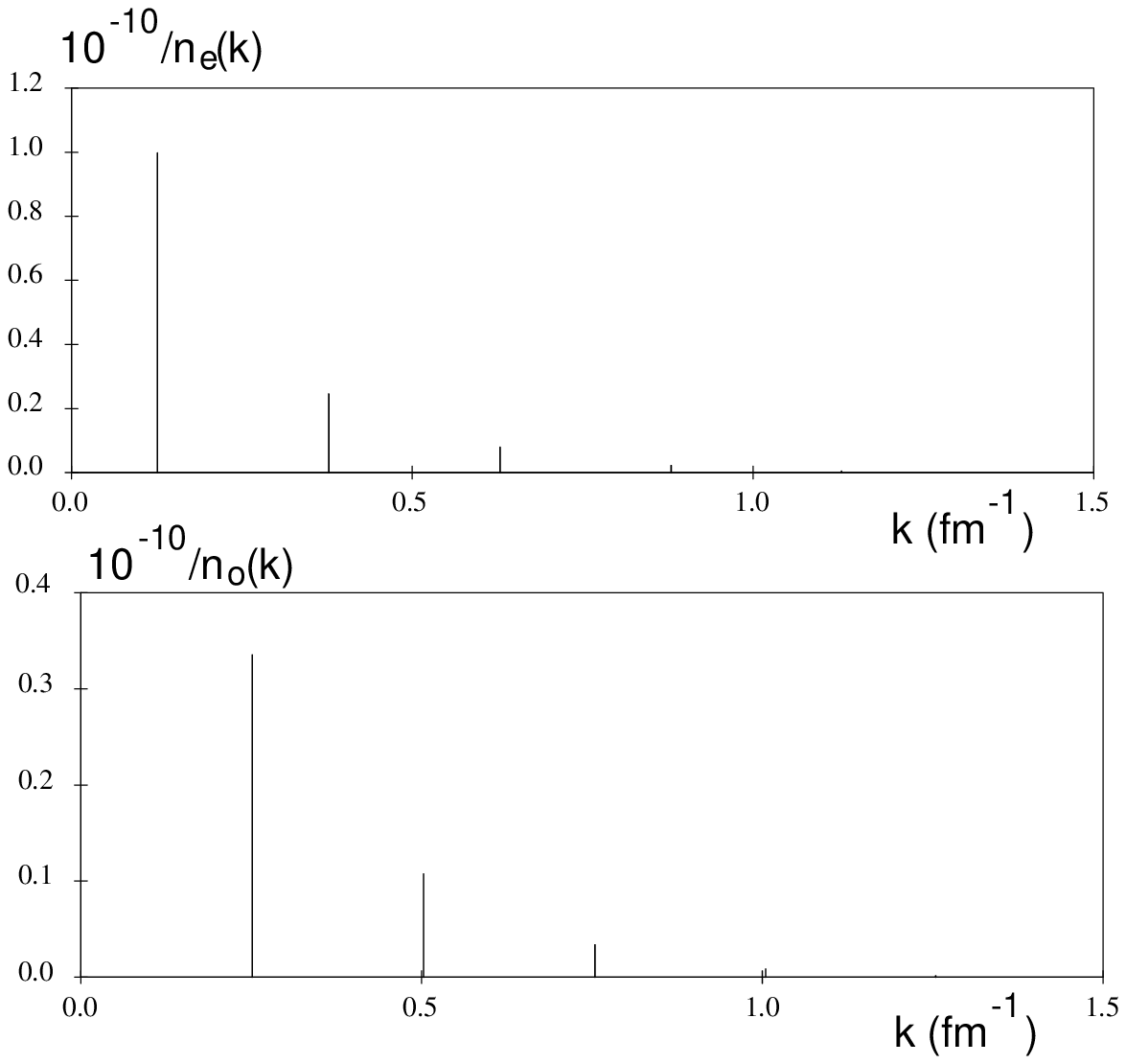}
	\caption{$ {10^{-10}}{n_e(k)}^{-1}$, and
${10^{-10}}{n_o(k)}^{-1}$ , as a function of {\sl k} in units of 
$fm^{-1}$ for a typical $\alpha$ decay barrier}
\label{fig1}
\end{figure}
The  poles in the complex momentum plane,
appear in pairs located symmetrically above and below the real 
momentum axis. The stationary wave functions of eqs.(\ref{even},\ref{odd})
consist of incoming and outgoing parts. 
The standard {\sl S} matrix, reflecting the behavior of the 
outgoing boundary condition at spatial infinity, has poles in the lower half  
of the complex momentum plane, whereas its inverse, 
corresponding to the incoming condition, has poles in the upper half.
Consequently, both sets of poles in the upper and lower complex momentum plane
appear when incoming and outgoing pieces are present.\ci{baz}.

As the poles are extremely sharp in position, it was necessary to use a
quadruple precision algorithm on an $\ds \alpha$ cluster mainframe to find them.
The imaginary parts of the poles are typically many orders of magnitude smaller 
than the real parts.

\noindent Figure 1 shows $\ds 10^{-10}{n_e(k)}^{-1}$, and
$\ds 10^{-10}{n_o(k)}^{-1}$ , for the parameters 
$\ds m=3727 MeV=18.88 fm^{-1}, x_0=12 fm, d=22 fm, 
\gamma=22 MeV = 0.11 fm^{-1}$, corresponding to a 
typical $\ds \alpha$ decay barrier.
The spikes are due to the extreme closeness of the minima of
the normalization factors to their complex zeros.

When eq.(\ref{expansion}) is evaluated by contour integration
in the complex $\ds k^2$ plane, the contour has to be closed
from below. In the lower half-plane the convergence is insured by 
the exponential $\ds e^{-i~\frac{k^2~t}{2~m}}.$
The poles are separated from each other and the integrand drops essentially to
zero outside the pole, hence, the wave function consists of an incoherent
sum of the residues of the different poles.
The negative imaginary part of each pole induces a time decaying exponential.
Each exponent determines a different decay constant and decay time.\ci{g1}
If the original wave function is even in space, as in
eq.(\ref{packet}), only {\sl even} poles show up in the unperturbed
wave function. 

The argument of the harmonic functions in eq.(\ref{ampl})
includes a potentially large denominator near the minima of the
normalization factors. The product of these even-odd factors near a minimum
is depicted in figures 2 and 3.
This product is not only finite, but of order one.
We were not able to produce an analytical formula for this product, 
although the numerical values are very suggestive.
\begin{figure}[htb]
\includegraphics[width=9cm,height=10cm]{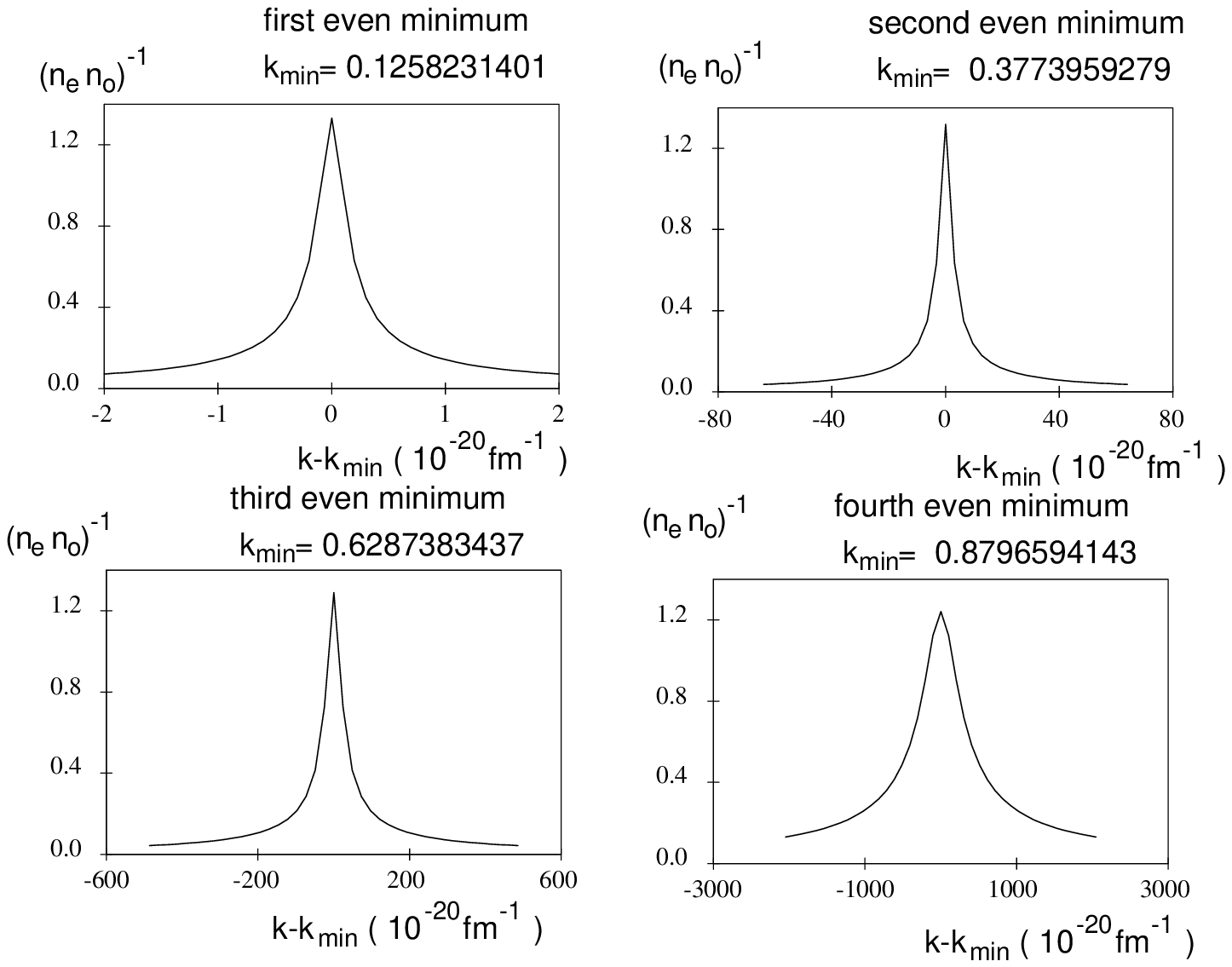}
\label{fig2}
		\caption{$ {n_o(k_{j,e})~n_{j,e}}^{-1}$, around the minima of
$n_e(k)$. Parameters as in figure 1}
\end{figure}

Figure 2 depicts $\ds ({n_e(k)~n_o(k)})^{-1}$, 
around the minima of $n_e(k)$. Note the degree of accuracy needed for the 
location of the minima. The parameters are those of figure 1.
Figure 3 depicts $\ds ({n_e(k)~n_o(k)})^{-1}$
around the minima of $n_o(k)$. 
\begin{figure}[htb]
\includegraphics[width=9cm,height=10cm]{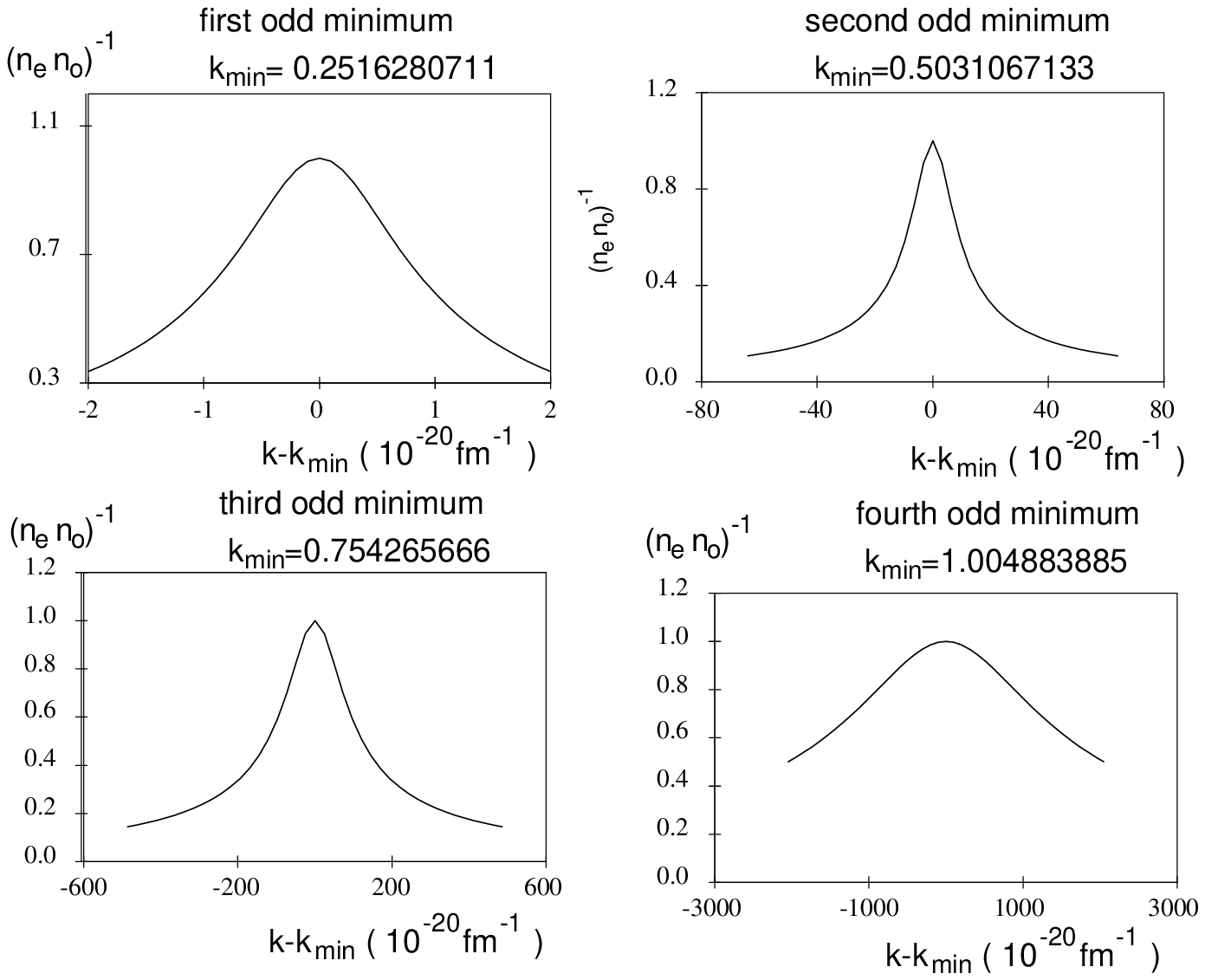}
\caption{$ {n_o(k_{j,e})~n_{j,e}}^{-1}$, around the minima of
$n_o(k)$. Parameters as in figure 1}
\label{fig3}
\end{figure}

Expanding the harmonic functions in eq.(\ref{ampl}), the exact expression
for the wave function in the region between the barriers becomes

\bea\label{psidev}
\Phi(x,t)&=&\Phi_{ee}(x,t)+\Phi_{oo}(x,t)+\Phi_{eo}(x,t)+\Phi_{oe}(x,t)\nono
\eea
\noindent where
\bea\label{fiee}
\Phi_{ee}(x,t)&=&\sum_j~\sum_{n=0}^{\infty}~G_j(x,t)\frac{(u_{j,e})^n}
{(n!)^2}\nono
G_j(x,t)&=&\frac{\pi^{3/4}\Delta~cos(s_{j,e}~x)~s_{j,e}~e^w}
{2~\sqrt{\beta_{j,e}\lambda_{j,e}}}\nono
u_{j,e}&=&-\bigg(\frac{\mu~
q_{j,e}~Y(t)}{2~m~n_o(q_{j,e})
\sqrt{\beta_{j,e}}}\bigg)^2\nono
w&=&-q_{j,e}(\Delta^2/4+i~t/(2~m))\nono
q_{j,e}&=&(k_{j,e})^2
-i\frac{\beta_{j,e}}{\lambda_{j,e}}\nono
s_{j,e}&=&\sqrt{q_{j,e}}
\eea
\vspace{1. true cm}
\bea\label{fioo}
\noindent \Phi_{oo}(x,t)&=&\sum_j\sum_{n=0}^{\infty}H_j(x,t)\frac{(u_{j,o})^n}{
(n!)^2~(2n+1)}\nono
H_j(x,t)&=&-\frac{\pi^{3/4}\Delta~sin(s_{j,o}~x)\mu~q_{j,o}~Y(t)~
e^w}{2~m~n_e(q_{j,o})^2\sqrt{\beta_{j,o}\lambda_{j,o}}}\nono
u_{j,o}&=&-\bigg(\frac{\mu~q_{j,o}~Y(t)}{2~m~n_e(q_{j,o})
~\sqrt{\beta_{j,o}}}\bigg)^2\nono
q_{j,o}&=&(k_{j,o})^2
-i\frac{\beta_{j,o}}{\lambda_{j,o}}\nono 
s_{j,o}&=&\sqrt{q_{j,o}}
\eea
\vspace{1. true cm}
\bea\label{fieo}
\noindent \Phi_{eo}(x,t)&=&\sum_j\sum_{n=1}^{\infty}L_j(x,t)\frac{(u_{j,o})^n}{
((n-1)!)^2~n~(2~n-1)}\nono
L_j(x,t)&=&\frac{2\pi^{3/4}\Delta~cos(s_{j,o}~x)~\sqrt{\beta_{j,o}}
~e^w}{{2~m~n_e(q_{j,o})^2
s_{j,o}\lambda_{j,o}}}\nono
\eea
\vspace{1. true cm}
\bea\label{fioe}
\noindent \Phi_{oe}(x,t)&=&\sum_j\sum_{n=0}^{\infty}M_j(x,t)\frac{(u_{j,e})^n}
{(2n+1)(n!)^2}\nono
M_j(x,t)&=&\frac{{\pi}^{3/4}\mu~Y(t)~s_{j,e}\Delta~cos(s_{j,e}~x)~
e^w}{2~m~n_o(q_{j,e})^2
\sqrt{\beta_{j,e}\lambda_{j,e}}}\nono
\eea
The series in eqs.(\ref{fiee},\ref{fioo},\ref{fieo},\ref{fioe}) 
can be expressed in terms of standard functions

\bea\label{fifi}
\Phi_{ee}(x,t)&=&\sum_j~G_j(x,t)~J_0(u_{j,e})\nono
\Phi_{oo}(x,t)&=&\sum_j~H_j(x,t)~\bigg[J_0(u_{j,o})-\frac{\pi}{2}\nono
&\bigg(&J_0(u_{j,o}){\bf H_1}(u_{j,o})-J_1(u_{j,o}){\bf H_0}(u_{j,o}
\bigg)\bigg]\nono
\Phi_{eo}(x,t)&=&\sum_j~L_j(x,t)\bigg[J_0(u_{j,o})-\frac{\pi}{2}\nono
&\bigg(&J_0(u_{j,o}){\bf H_1}(u_{j,o})
-J_1(u_{j,o}){\bf  H_0}(u_{j,o})\bigg)\bigg]\nono
\Phi_{oe}(x,t)&=&\sum_j~M_j(x,t)\bigg[2~J_0(u_{j,e})-\frac{2~J_1(u_{j,e})}
{u_{j,e}}-\pi\bigg(\nono
&&J_0(u_{j,e}){\bf H_1}(u_{j,e})-J_1(u_{j,e}){\bf H_0}(u_{j,e})
\bigg)\bigg]
\eea

\noindent where $\ds J_{0,1}$ denotes the Bessel function of 
the order (0,1), and $\ds {\bf H_{0,1}}$ is the Struve {\bf H} function.
\ci{abram} 

For long times we can use the asymptotic expansions of the Bessel and
Struve functions to obtain\ci{abram}

\bea\label{asympfi}
\Phi_{ee}(x,t)&\ra&\sum_j~G_j(x,t)~\bigg(
\sqrt{\frac{2}{u_{j,e}~\pi}}sin(u_{j,e}\nono
&+&\frac{\pi}{4})+O(u_{j,e}^{-\frac{3}{2}})\bigg)\nono
\Phi_{oo}(x,t)&\ra&\sum_j~H_j(x,t)~\bigg(\frac{1}{u_{j,o}}
+O(u_{j,o}^{-\frac{3}{2}})\bigg)\nono
\Phi_{eo}(x,t)&\ra&\sum_j~L_j(x,t)~\bigg(\frac{1}{u_{j,o}}
+O(u_{j,o}^{-\frac{3}{2}})\bigg)\nono
\Phi_{oe}(x,t)&\ra&\sum_j~M_j(x,t)~\bigg(\frac{2}{u_{j,e}}
+O(u_{j,e}^{-\frac{3}{2}})\bigg)
\eea

\noindent From eq.(\ref{asympfi}) it is clear that for long times the even-even
component is essentially the dominant one. 
\begin{figure}[htb]
\includegraphics[width=9cm,height=9cm]{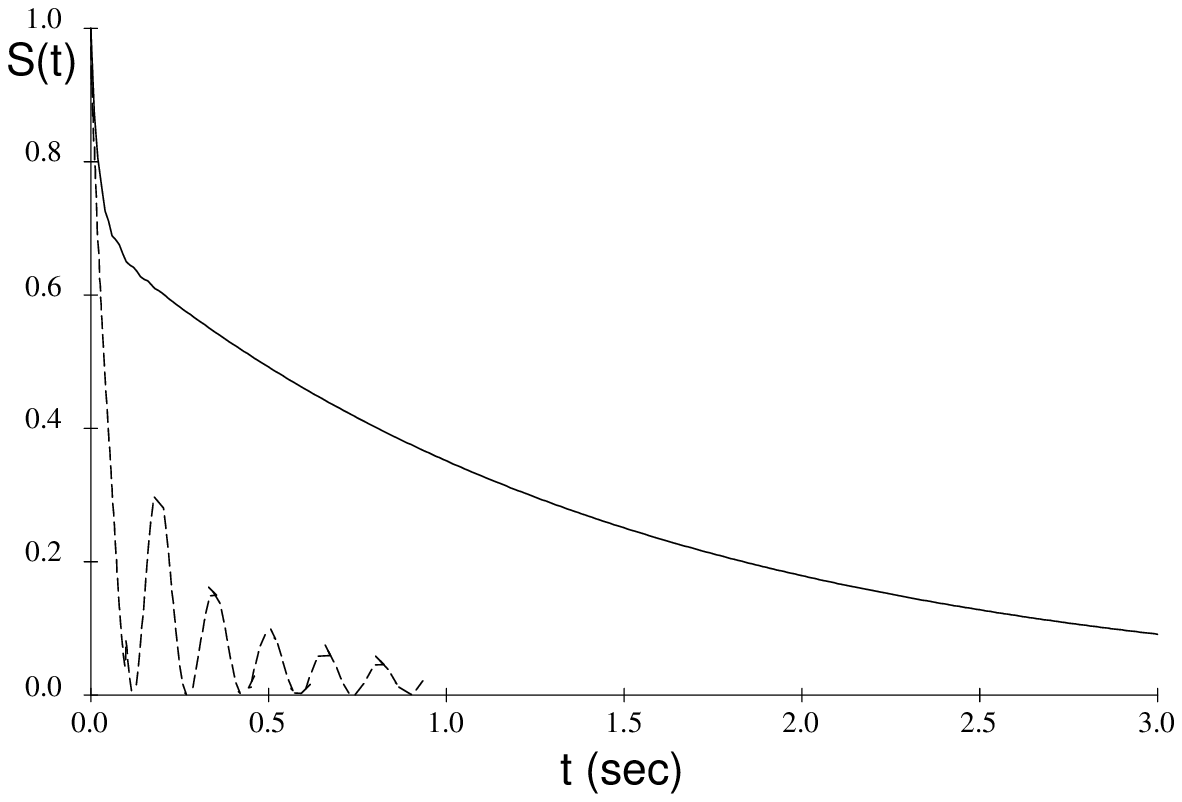}
\caption{S(t) of eq.(\ref{surv}), without, full line, and with 
harmonic perturbation, dashed line, see text for parameters}
\label{fig4}
\end{figure}
\begin{figure}[htb]
\includegraphics[width=9cm,height=9cm]{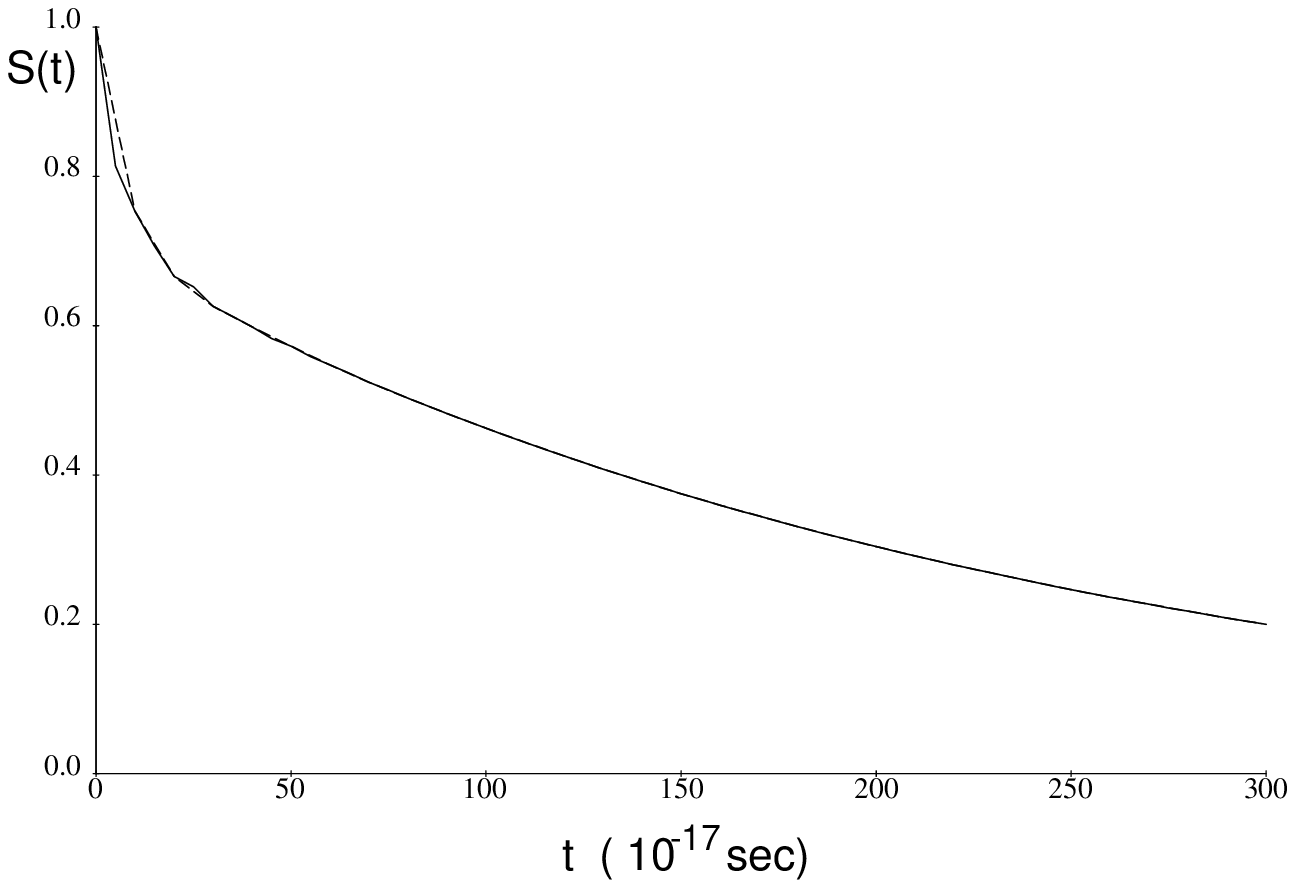}
\caption{S(t) of eq.(\ref{surv}), without, full line, and with 
harmonic perturbation, dashed line, for $\delta$ barriers}
	\label{fig5}
\end{figure}
We can now evaluate the influence of the perturbation on the decay time.
Figure 4 depicts the nonescape probability of the system in the region
between the barriers

\bea\label{surv}
S(t)=\int_{-x_0}^{x_0}~|\Psi(x,t)|^2~dx
\eea
 
\noindent for the square barrier case, we have used the parameters relevant for
$\ds \alpha$ decay mentioned below eq.(\ref{h1}) and an intial wave packet
of width $\ds \Delta=4 fm$. The lower curve corresponds
to (see eq.(\ref{th})) an electric field of $\ds E=0.1 Volt/m$, and a 
frequency of $\ds \omega=10\pi~10^{17} Hz$. The aim of this work is to present the
analytical solutions and show the acceleration of the decay process. 
A more realistic choice of parameters demands a thorough numerical
evaluation of the integrals outside the poles and 
is postponed for future work.
For a time harmonic perturbation(\ref{th}), 
there are two major corrections to the unperturbed decay
rate. The amplitude of the wave function is to lowest order multiplied by a 
factor of the form $\ds (1-ct^2/4)$, with $\ds c\approx \frac{\mu^2~k^2}
{m^2\omega^2}$ from the series in $\ds u_{j,e}$. This factor depletes 
the wave at times of the order of the
period of the harmonic perturbation for moderate values of $\mu$, much faster 
than the actual decay time determined by the poles. A second  
correction already noted in\ci{g1}, arises when considering 
the higher order poles $\ds j>1$ in the sum of eq.(\ref{fiee}).
The latter generates a different decay time for each pole\ci{g1}.

Figure 5 shows the results for delta function barriers 
$\ds \nu~\delta(x-x_0)+\nu\delta(x+x_0)$, with parameters
corresponding to the square barrier case $\ds \nu=\gamma(d-x_0)$. 
The unperturbed
decay time of around $\ds~10^{-16}sec$ is orders of magnitude smaller 
than the square barrier one. The assisted tunneling is
barely noticeable.

In summary, we presented a simple analytical method to calculate 
assisted tunneling.
The technique can be generalized to three dimensions, and to other potentials
and initial metastable states. Such endeavor is worth pursuing due to the
dramatic impact assisted tunneling may have on nuclear technologies as well as 
on solid state devices.
The question of electron screening for the nuclear decay acceleration must also
be addressed to produce experimentally meaningful predictions.

\begin{acknowledgments}
\noindent It is a pleasure to acknowledge the anonymous referee's 
remarks and corrections.
\end{acknowledgments}

\section{\sl Appendix A: Perturbation matrix element}

The matrix element of eq.(\ref{superp}) does not vanish for transitions
between even and odd states solely.
The matrix element evidently diverges and we seek to evaluate the divergent 
part. The most important contribution comes from the outer region of 
$\ds |x|>d$. In the inner region there will be a finite contribution that
is negligible as compared to the $\ds \delta$ 
function contribution of the outer region.
 Consequently we can use the outer region expression throughout.

The functions $\ds \chi_{e,o}$ are then replaced for all {\sl x} by

\bea\label{eo}
\sqrt{\pi}~n_e(k)\phi_e(k)&=&(E~e^{i~kx}+ E^*~e^{-i~k x}),~x>0\nono
\sqrt{\pi}~n_e(k)\phi_e(k)&=&(E~e^{-i~kx}+ E^*~e^{i~k x}),~x<0\nono
\sqrt{\pi}~n_o(k)\phi_o(k)&=&(F~e^{i~kx}+F^*~e^{-i~k x}),x>0\nono
\sqrt{\pi}~n_o(k)\phi_o(k)&=&-(F~e^{-i~kx}+F^*~e^{i~k x}),x<0\nono
E&=&\frac{C_1-i~D_1}{2}\nono
F&=&\frac{C_2-i~D_2}{2}
\eea

\noindent Taking advantadge of the symmetry properties of the wave functions, 
the matrix element of eq.(\ref{superp}) becomes

\bea\label{superp1}
M_{k',k}&=&\int_{-\infty}^{\infty}~e^{-\frac{i(k'^2-k^2)~t}{2m}}
\chi_e(k,x)\frac{\dd \chi_o(k',x)}{\dd x} dx\nono
&=&\frac{2}{\pi~n_e(k)~n_o(k')}\sum_{i=1}^4~(I_i(x\ra\infty)
-I_i(x=0))\nono
I_1(x)&=&i~k'E\tilde{F}~\frac{e^{i(k+k')x}}{i~(k+k')}\nono
I_2(x)&=&-i~k'E\tilde{F}^*~\frac{e^{i(k-k')x}}{i~(k-k')}\nono
I_3(x)&=&i~k'E^*\tilde{F}~\frac{e^{i(-k+k')x}}{i~(-k+k')}\nono
I_4(x)&=&-i~k'E^*\tilde{F}^*\frac{e^{-i(k+k')x}}{-i~(k+k')}
\eea

\noindent where the {\sl tilde} denotes a factor that depends on {\sl k'}
instead of {\sl k}. The all important contribution comes from $\ds k'\to k$

\bea\label{q}
M_{k',k}&\approx& 2~i~k'(E^*~F-E~F^*)\frac{\delta(k-k')}{n_e(k)n_o(k')}
\eea

\noindent A lengthy but straightforward calculation then gives the result
of eq.(\ref{superp})
\bea\label{superp2}
M_{k',k}=\frac{k}{n_e(k)~n_o(k)}\delta(k-k')
\eea

\end{document}